\documentclass[prd,reprint,superscriptaddress,amsmath,amssymb,aps]{revtex4-2}
\usepackage{xcolor}
\usepackage{graphicx}
\usepackage{dcolumn}
\usepackage{newtxtext}
\usepackage{bm}
\definecolor{APSblue}{RGB}{46,49,145}
\usepackage[ colorlinks=true, urlcolor=APSblue, citecolor=APSblue, linkcolor=black ]{hyperref}
\usepackage{multirow} 
\usepackage{orcidlink}
\usepackage{siunitx}
\usepackage{enumitem}
\usepackage{placeins} 
\usepackage{tablefootnote}
\usepackage{soul}
\usepackage{textcomp}
\usepackage{verbatim}
\begin{document}
\title{\texorpdfstring{Search for nucleon decay via $\bm{p\rightarrow\nu\pi^{+}}$ and $\bm{n\rightarrow\nu\pi^{0}}$ in 0.484\,Mton-year of Super-Kamiokande data}{Search for nucleon decay via p -> nu pi+ and n -> nu pi0 in 0.484 Mton-year of Super-Kamiokande data}}
\newcommand{\AFFicrr}{\affiliation{Kamioka Observatory, Institute for Cosmic Ray Research, University of Tokyo, Kamioka, Gifu 506-1205, Japan}}
\newcommand{\AFFkashiwa}{\affiliation{Research Center for Cosmic Neutrinos, Institute for Cosmic Ray Research, University of Tokyo, Kashiwa, Chiba 277-8582, Japan}}
\newcommand{\AFFipmu}{\affiliation{Kavli Institute for the Physics and
Mathematics of the Universe (WPI), The University of Tokyo Institutes for Advanced Study,
University of Tokyo, Kashiwa, Chiba 277-8583, Japan }}
\newcommand{\AFFmad}{\affiliation{Department of Theoretical Physics, University Autonoma Madrid, 28049 Madrid, Spain}}
\newcommand{\AFFubc}{\affiliation{Department of Physics and Astronomy, University of British Columbia, Vancouver, BC, V6T1Z4, Canada}}
\newcommand{\AFFbu}{\affiliation{Department of Physics, Boston University, Boston, MA 02215, USA}}
\newcommand{\AFFuci}{\affiliation{Department of Physics and Astronomy, University of California, Irvine, Irvine, CA 92697-4575, USA }}
\newcommand{\AFFcsu}{\affiliation{Department of Physics, California State University, Dominguez Hills, Carson, CA 90747, USA}}
\newcommand{\AFFcnm}{\affiliation{Institute for Universe and Elementary Particles, Chonnam National University, Gwangju 61186, Korea}}
\newcommand{\AFFduke}{\affiliation{Department of Physics, Duke University, Durham NC 27708, USA}}
\newcommand{\AFFgifu}{\affiliation{Department of Physics, Gifu University, Gifu, Gifu 501-1193, Japan}}
\newcommand{\AFFgist}{\affiliation{GIST College, Gwangju Institute of Science and Technology, Gwangju 500-712, Korea}}
\newcommand{\AFFuh}{\affiliation{Department of Physics and Astronomy, University of Hawaii, Honolulu, HI 96822, USA}}
\newcommand{\AFFicl}{\affiliation{Department of Physics, Imperial College London , London, SW7 2AZ, United Kingdom }}
\newcommand{\AFFkek}{\affiliation{High Energy Accelerator Research Organization (KEK), Tsukuba, Ibaraki 305-0801, Japan }}
\newcommand{\AFFkobe}{\affiliation{Department of Physics, Kobe University, Kobe, Hyogo 657-8501, Japan}}
\newcommand{\AFFkyoto}{\affiliation{Department of Physics, Kyoto University, Kyoto, Kyoto 606-8502, Japan}}
\newcommand{\AFFliv}{\affiliation{Department of Physics, University of Liverpool, Liverpool, L69 7ZE, United Kingdom}}
\newcommand{\AFFmiyagi}{\affiliation{Department of Physics, Miyagi University of Education, Sendai, Miyagi 980-0845, Japan}}
\newcommand{\AFFnagoya}{\affiliation{Institute for Space-Earth Environmental Research, Nagoya University, Nagoya, Aichi 464-8602, Japan}}
\newcommand{\AFFkmi}{\affiliation{Kobayashi-Maskawa Institute for the Origin of Particles and the Universe, Nagoya University, Nagoya, Aichi 464-8602, Japan}}
\newcommand{\AFFpol}{\affiliation{National Centre For Nuclear Research, 02-093 Warsaw, Poland}}
\newcommand{\AFFsuny}{\affiliation{Department of Physics and Astronomy, State University of New York at Stony Brook, NY 11794-3800, USA}}
\newcommand{\AFFokayama}{\affiliation{Department of Physics, Okayama University, Okayama, Okayama 700-8530, Japan }}
\newcommand{\AFFosaka}{\affiliation{Department of Physics, Osaka University, Toyonaka, Osaka 560-0043, Japan}}
\newcommand{\AFFox}{\affiliation{Department of Physics, Oxford University, Oxford, OX1 3PU, United Kingdom}}
\newcommand{\AFFqmul}{\affiliation{School of Physics and Astronomy, Queen Mary University of London, London, E1 4NS, United Kingdom}}
\newcommand{\AFFregina}{\affiliation{Department of Physics, University of Regina, 3737 Wascana Parkway, Regina, SK, S4SOA2, Canada}}
\newcommand{\AFFseoul}{\affiliation{Department of Physics and Astronomy, Seoul National University, Seoul 151-742, Korea}}
\newcommand{\AFFsheff}{\affiliation{School of Mathematical and Physical Sciences, University of Sheffield, S3 7RH, Sheffield, United Kingdom}}
\newcommand{\AFFshizuokasc}{\affiliation{Department of Informatics in
Social Welfare, Shizuoka University of Welfare, Yaizu, Shizuoka, 425-8611, Japan}}
\newcommand{\AFFstfc}{\affiliation{STFC, Rutherford Appleton Laboratory, Harwell Oxford, and Daresbury Laboratory, Warrington, OX11 0QX, United Kingdom}}
\newcommand{\AFFskk}{\affiliation{Department of Physics, Sungkyunkwan University, Suwon 440-746, Korea}}
\newcommand{\AFFtodai}{\affiliation{Department of Physics, University of Tokyo, Bunkyo, Tokyo 113-0033, Japan }}
\newcommand{\AFFtit}{\affiliation{Department of Physics, Institute of Science Tokyo, Meguro, Tokyo 152-8551, Japan }}
\newcommand{\AFFtus}{\affiliation{Department of Physics and Astronomy, Faculty of Science and Technology, Tokyo University of Science, Noda, Chiba 278-8510, Japan }}
\newcommand{\AFFtriumf}{\affiliation{TRIUMF, 4004 Wesbrook Mall, Vancouver, BC, V6T2A3, Canada }}
\newcommand{\AFFtokai}{\affiliation{Department of Physics, Tokai University, Hiratsuka, Kanagawa 259-1292, Japan}}
\newcommand{\AFFtsinghua}{\affiliation{Department of Engineering Physics, Tsinghua University, Beijing, 100084, China}}
\newcommand{\AFFynu}{\affiliation{Department of Physics, Yokohama National University, Yokohama, Kanagawa, 240-8501, Japan}}
\newcommand{\AFFllr}{\affiliation{Ecole Polytechnique, IN2P3-CNRS, Laboratoire Leprince-Ringuet, F-91120 Palaiseau, France }}
\newcommand{\AFFbari}{\affiliation{ Dipartimento Interuniversitario di Fisica, INFN Sezione di Bari and Universit\`a e Politecnico di Bari, I-70125, Bari, Italy}}
\newcommand{\AFFnapoli}{\affiliation{Dipartimento di Fisica, INFN Sezione di Napoli and Universit\`a di Napoli, I-80126, Napoli, Italy}}
\newcommand{\AFFroma}{\affiliation{INFN Sezione di Roma and Universit\`a di Roma ``La Sapienza'', I-00185, Roma, Italy}}
\newcommand{\AFFpadova}{\affiliation{Dipartimento di Fisica, INFN Sezione di Padova and Universit\`a di Padova, I-35131, Padova, Italy}}
\newcommand{\AFFkeio}{\affiliation{Department of Physics, Keio University, Yokohama, Kanagawa, 223-8522, Japan}}
\newcommand{\AFFwinnipeg}{\affiliation{Department of Physics, University of Winnipeg, MB R3J 3L8, Canada }}
\newcommand{\AFFkcl}{\affiliation{Department of Physics, King's College London, London, WC2R 2LS, UK }}
\newcommand{\AFFwarwick}{\affiliation{Department of Physics, University of Warwick, Coventry, CV4 7AL, UK }}
\newcommand{\AFFral}{\affiliation{Rutherford Appleton Laboratory, Harwell, Oxford, OX11 0QX, UK }}
\newcommand{\AFFwu}{\affiliation{Faculty of Physics, University of Warsaw, Warsaw, 02-093, Poland }}
\newcommand{\AFFbcit}{\affiliation{Department of Physics, British Columbia Institute of Technology, Burnaby, BC, V5G 3H2, Canada }}
\newcommand{\AFFtohoku}{\affiliation{Department of Physics, Faculty of Science, Tohoku University, Sendai, Miyagi, 980-8578, Japan }}
\newcommand{\AFFicise}{\affiliation{Institute For Interdisciplinary Research in Science and Education, ICISE, Quy Nhon, 55121, Vietnam }}
\newcommand{\AFFilance}{\affiliation{ILANCE, CNRS - University of Tokyo International Research Laboratory, Kashiwa, Chiba 277-8582, Japan}}
\newcommand{\AFFibs}{\affiliation{Center for Underground Physics, Institute for Basic Science (IBS), Daejeon, 34126, Korea}}
\newcommand{\AFFglasgow}{\affiliation{School of Physics and Astronomy, University of Glasgow, Glasgow, Scotland, G12 8QQ, United Kingdom}}
\newcommand{\AFFoecu}{\affiliation{Media Communication Center, Osaka Electro-Communication University, Neyagawa, Osaka, 572-8530, Japan}}
\newcommand{\AFFminn}{\affiliation{School of Physics and Astronomy, University of Minnesota, Minneapolis, MN  55455, USA}}
\newcommand{\AFFsilesia}{\affiliation{August Che\l{}kowski Institute of Physics, University of Silesia in Katowice, 75 Pu\l{}ku Piechoty 1, 41-500 Chorz\'{o}w, Poland}}
\newcommand{\AFFtoyama}{\affiliation{Faculty of Science, University of Toyama, Toyama City, Toyama 930-8555, Japan}}
\newcommand{\AFFbmcc}{\affiliation{Science Department, Borough of Manhattan Community College / City University of New York, New York, New York, 1007, USA.}}
\newcommand{\AFFnumazu}{\affiliation{National Institute of Technology, Numazu College, Numazu, Shizuoka 410-8501, Japan}}

\AFFicrr
\AFFkashiwa
\AFFmad
\AFFbmcc
\AFFbu
\AFFbcit
\AFFuci
\AFFcsu
\AFFcnm
\AFFduke
\AFFllr
\AFFgifu
\AFFgist
\AFFglasgow
\AFFuh
\AFFibs
\AFFicise
\AFFicl
\AFFbari
\AFFnapoli
\AFFpadova
\AFFroma
\AFFilance
\AFFkeio
\AFFkek
\AFFkcl
\AFFkobe
\AFFkyoto
\AFFliv
\AFFminn
\AFFmiyagi
\AFFnagoya
\AFFkmi
\AFFpol
\AFFnumazu
\AFFsuny
\AFFokayama
\AFFoecu
\AFFox
\AFFral
\AFFseoul
\AFFsheff
\AFFshizuokasc
\AFFsilesia
\AFFstfc
\AFFskk
\AFFtohoku
\AFFtodai
\AFFipmu
\AFFtit
\AFFtus
\AFFtoyama
\AFFtriumf
\AFFtsinghua
\AFFwu
\AFFwarwick
\AFFwinnipeg
\AFFynu

\author{S.~Jung\orcidlink{0009-0007-8244-8106}}
\AFFseoul


\author{K.~Abe\orcidlink{0009-0000-9620-788X}}
\AFFicrr
\AFFipmu
\author{Y.~Asaoka\orcidlink{0000-0001-6440-933X}}
\AFFicrr
\AFFipmu
\author{M.~Harada\orcidlink{0000-0003-3273-946X}}
\AFFicrr
\author{Y.~Hayato\orcidlink{0000-0002-8683-5038}}
\AFFicrr
\AFFipmu
\author{K.~Hiraide\orcidlink{0000-0003-1229-9452}}
\AFFicrr
\AFFipmu
\author{T.~H.~Hung}
\AFFicrr
\author{K.~Ieki\orcidlink{0000-0002-7791-5044}}
\author{M.~Ikeda\orcidlink{0000-0002-4177-5828}}
\AFFicrr
\AFFipmu
\author{J.~Kameda}
\AFFicrr
\AFFipmu
% ...


\collaboration{The Super-Kamiokande Collaboration}
\noaffiliation

\date{\today}
\begin{abstract}
We present the results of searches for nucleon decays via $p\rightarrow\nu\pi^{+}$ and $n\rightarrow\nu\pi^{0}$ using a 0.484\,Mt$\cdot$yr exposure of Super-Kamiokande I-V data covering the entire pure water phase of the experiment. Various improvements on the previous 2014 nucleon decay search~[\href{https://doi.org/10.1103/PhysRevLett.113.121802}{Phys.\ Rev.\ Lett.\ \textbf{113}, 121802 (2014)}], which used an exposure of 0.173\,Mt$\cdot$yr, are incorporated. The physics models related to pion production and nuclear interaction are refined with external data, and a more comprehensive set of systematic uncertainties, now including those associated with the atmospheric neutrino flux and pion production channels is considered. Also, the fiducial volume has been expanded by 21\%. No significant indication of a nucleon decay signal is found beyond the expected background. Lower bounds on the nucleon partial lifetimes are determined to be $3.5\times10^{32}$\,yr for $p\rightarrow\nu\pi^{+}$ and $1.4\times10^{33}$\,yr for $n\rightarrow\nu\pi^{0}$ at 90\% confidence level.
\end{abstract}
\maketitle
\section{INTRODUCTION}
Grand unified theories (GUTs) are proposed featuring extended gauge symmetry groups for the unification of electromagnetic, weak, and strong interactions of the Standard Model (SM)~\cite{PhysRevLett.32.438,PhysRevD.10.275,FRITZSCH1975193}. In addition to unifying the three fundamental forces, they are motivated by their ability to provide hints for several outstanding questions that the SM does not address, such as the quantization of electric charge, prediction of free parameters in the SM, and the matter-antimatter asymmetry of the Universe via baryogenesis \cite{PhysRevD.103.043504}. In contrast to the SM, GUTs represent leptons and quarks within the same multiplets, allowing baryon-number-violating nucleon decays. The typical energy scale of unification is expected to exceed $10^{15}$\,GeV, far beyond the energy scale of the future colliders. Nucleon decay searches offer a direct probe for the viability of various SM extensions. With a large target mass of 50 kt, the Super-Kamiokande (SK) water Cherenkov detector is ideal for performing such searches. 

While simple GUT models, based on the minimal $SU(5)$~\cite{LANGACKER1981185,DIMOPOULOS1981150,SAKAI19810601,ELLIS198243,DIMOPOULOS1982133} have been ruled out by experiments~\cite{PhysRevLett.51.27,HIRATA1989308,PhysRevLett.81.3319,PhysRevD.65.055009}, more elaborate models have been suggested, such as $SU(5)$ with flipped fermion assignments~\cite{DERENDINGER1984170,BARR1982219} which solves the Higgs splitting problem~\cite{ANTONIADIS1987231}, or for a range of $SO(10)$ scenarios which explain the fermion masses and mixing parameters~\cite{PhysRevLett.70.2845,PhysRevD.64.053015,Matsuda2002SO10,TakeshiFukuyama2002,GOH2004105}, with or without supersymmetry. Among these models, the nucleon decays via $p\rightarrow\nu\pi^{+}$ and $n\rightarrow\nu\pi^{0}$ are predicted with comparable decay widths alongside $p\rightarrow e^{+}\pi^{0}$~\cite{ELLIS19881,Ellis2020} and dominantly in certain regions of parameter space in a minimal supersymmetry $SO(10)$ with baryon-number-minus-lepton-number symmetry broken by a 126-dimensional Higgs field~\cite{GOH2004105}.

Several experiments have searched for these decay modes \cite{HIRATA1989308,PhysRevLett.113.121802,PhysRevD.59.052004,PhysRevD.62.092003}, but no clear signal has been observed to date. The most stringent lower limits were set by SK, which are $3.9\times 10^{32}$\,yr for $p\rightarrow\nu\pi^{+}$ and $1.1\times 10^{33}$\,yr for $n\rightarrow\nu\pi^{0}$ at 90\% confidence level~\cite{PhysRevLett.113.121802}. In this study, we improve the previous results with the following: (i) With the addition of SK IV-V data to SK I-III, the detector live time has increased by 132\% to 17.8\,yr. (ii) The fiducial volume of SK is enlarged by 21\% using an improved analysis technique. (iii) Physics models related to pion production and nuclear interaction are updated in recent studies with external data~\cite{PhysRevD.95.012004,PhysRevD.102.112011}. (iv) Additional systematic uncertainties associated with the atmospheric neutrino flux and neutrino-nucleon interaction models have been newly implemented.
\section{SUPER-KAMIOKANDE}\label{section:superk}
SK is a cylindrical water Cherenkov detector with a diameter of 39.3\,m and a height of 41.4\,m, filled with 50\,kt of ultrapure water. The detector is located beneath the peak of Mount Ikeno with 1000\,m of rock overburden, equivalent to 2700\,m of water. The inner detector (ID) and the outer detector (OD) volumes are optically separated. The ID is viewed by 50-cm-diameter photomultiplier tubes (PMTs), while the OD is equipped with 20-cm-diameter PMTs with acrylic wavelength shifting plates. Details of the SK detector are described in Ref.~\cite{FUKUDA2003418}.

The SK detector operations, corresponding to the data used in this analysis, are classified into five detector phases, denoted by SK-I (1489.2 days), SK-II (798.6 days), SK-III (518.1 days), SK-IV (3244.4 days), and SK-V (461.0 days). SK-I started with the 11146 ID PMTs providing 40\% photo coverage and the 1885 OD PMTs. However, in November 2001, more than half of the PMTs were destroyed due to a chain-reaction implosion inside the tank. After the accident, the remaining PMTs were rearranged, and SK-II began with a reduced 19\% ID photo coverage. The ID PMTs have been covered by fiber-reinforced plastic to prevent the chain-reaction implosion since the accident. Following a full reconstruction of the ID PMTs, SK-III began, restoring the 40\% ID photocathode coverage. From SK-IV, front-end electronics were updated to achieve continuous recording of all the PMT hit information for the software triggering and wide charge dynamic range~\cite{NISHINO2009710}. The upgraded system has improved the tagging efficiency for secondary particles such as Michel electrons after muon decay. SK-IV ended with the installation of a new water circulation system and the replacement of dead PMTs. Subsequently, SK-V continued to take pure water data and ended with the dissolution of gadolinium into the water.

The detector is calibrated using controlled data samples to ensure precise and consistent measurement of physics quantities. We measure absorption and scattering coefficients of optical photons, as well as the reflectivity of the PMT surface, using a collimated laser beam. For the analysis, particle identification based on the Cherenkov hit pattern relies on accurate calibration of optical photon tracking. The energy scale is calibrated using various natural sources, including cosmic-ray-stopping muons (sub-GeV and multi-GeV), decay electrons from stopping muons (tens of MeV), and neutral pions produced in atmospheric neutrino interactions via weak neutral current (hundreds of MeV). These calibrations are used to assign energy-scale uncertainties in this analysis. Details of the detector calibrations for each SK phase are described in Refs.~\cite{FUKUDA2003418,ABE2014253,doi:10.1142/9789819801107_0008,PhysRevD.109.072014}.
\begin{figure*}[!htbp]
	\includegraphics[width=0.32\textwidth]{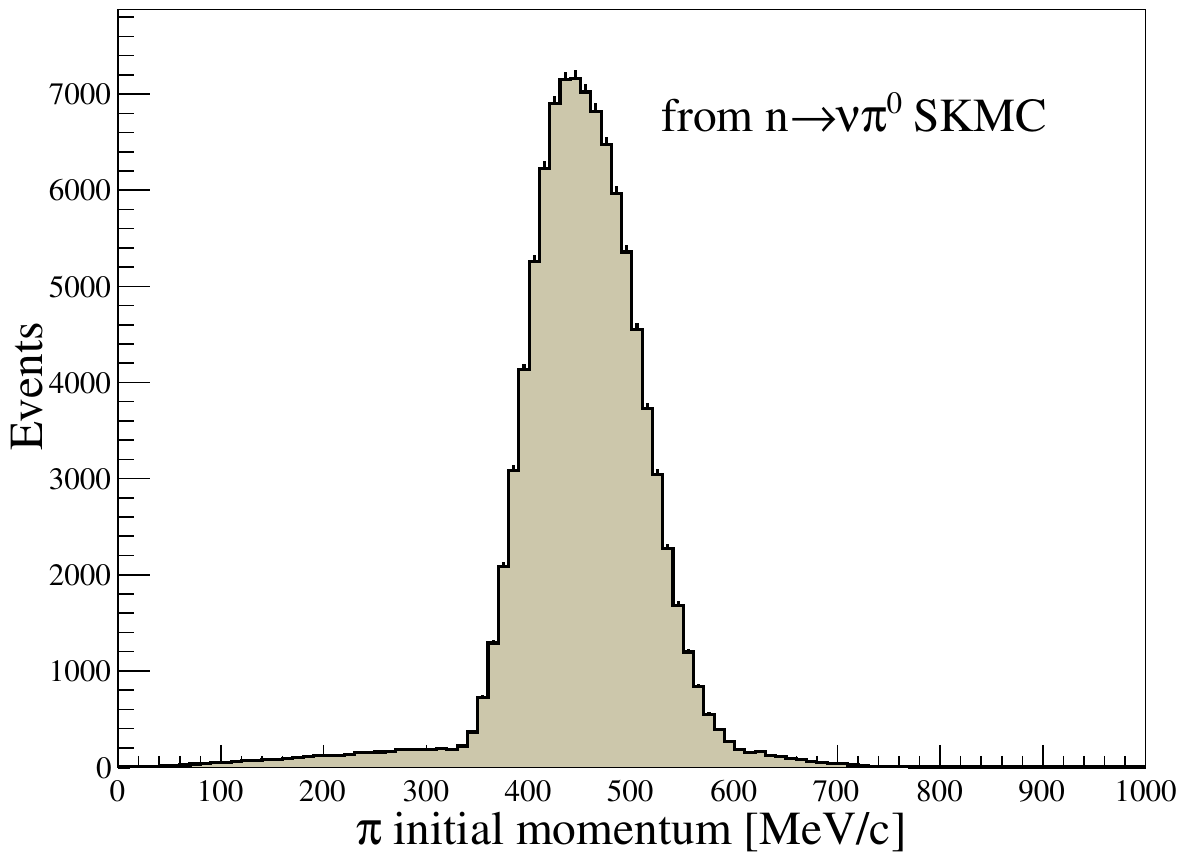}
	\includegraphics[width=0.32\textwidth]{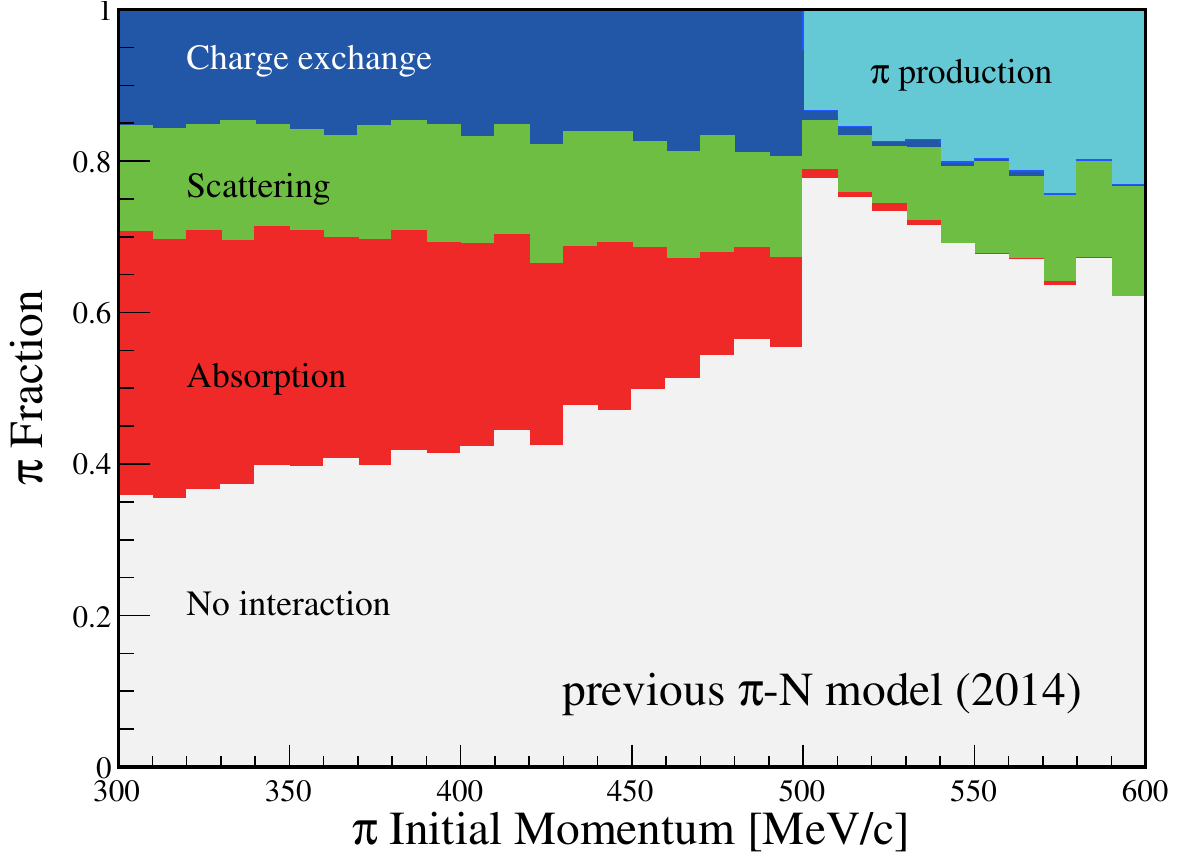}
	\includegraphics[width=0.32\textwidth]{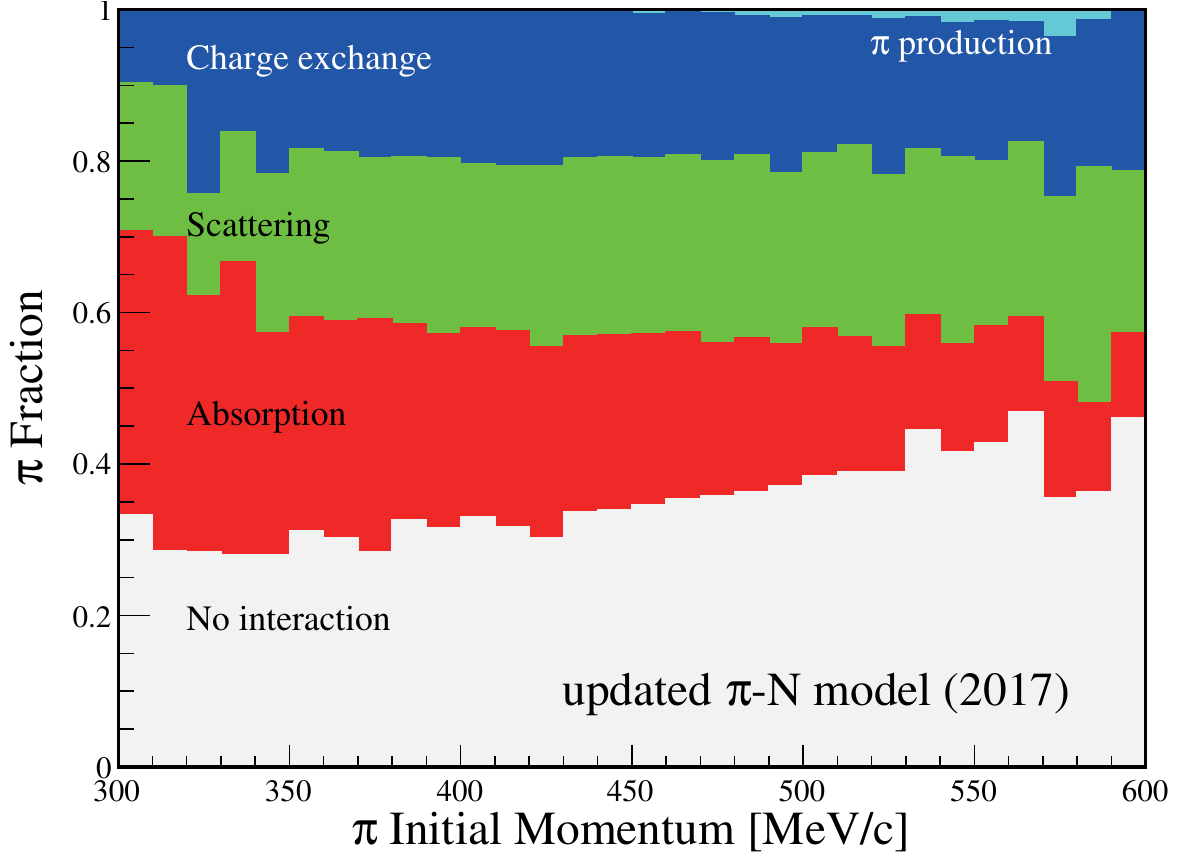}
	\caption{\label{figure:picrs}The left panel shows true neutral pion initial momentum from a $n\rightarrow\nu\pi^0$ SK MC study. Center panel shows cumulative fractions of nuclear interactions for neutral pion as a function of its true momentum predicted by the previous model~\cite{PhysRevLett.113.121802}. The sharp change in fractions at $\sim$500\,MeV/$c$ indicates a lack of experimental data points above this momentum. The right panel shows the updated model based on the additional experimental data~\cite{PhysRevD.95.012004}. The fraction for each nuclear interaction is labeled at the corresponding color-filled region: ``No interaction" means the neutral pion escapes the nucleus without nuclear interaction,  ``Absorption" is denoted for pion absorption, ``scattering" indicates pion scattering events, ``charge exchange" is the case where the pion charge is exchanged, and ``$\pi$ production" is the production of pions including multiple hadrons.}
\end{figure*}
\section{SIMULATION}\label{section:simulation}
Nucleon decay signals are generated using Monte Carlo (MC) simulation. Signal events for $p\rightarrow\nu\pi^{+}$ are generated from protons of hydrogen or oxygen, while events for $n\rightarrow\nu\pi^{0}$ are generated from neutrons of oxygen in water molecules. A proton in hydrogen is treated as a stationary particle with a mass equal to the proton's rest mass. In contrast, eight protons and neutrons in an oxygen nucleus are treated as bound particles whose momenta and masses are determined by Fermi motion and nuclear binding energy. Based on the nuclear shell model~\cite{mayer1955elementary}, an initial bound nucleon state is assigned as either an \textit{s} state (25\%) or a \textit{p} state (75\%). For bound nucleons, Fermi momentum is simulated based on the proton spectral function measured by the electron-carbon scattering experiment~\cite{NAKAMURA1976381}, and the effective mass is calculated by subtracting the nuclear binding energy from the nucleon rest mass. The nuclear binding energy is simulated using normalized Gaussian distributions: The \textit{s}-state distribution has a mean of 39.0\,MeV with a standard deviation of 10.2\,MeV, and the \textit{p}-state distribution has a mean of 15.5\,MeV with a standard deviation of 3.8\,MeV. Considering the nuclear medium effects, 10\% of bound nucleon decays are treated as correlated decays, where the decay kinematics are broadened due to surrounding nucleons~\cite{YAMAZAKI19991}.

Atmospheric neutrino interactions with nucleons are backgrounds for the nucleon decay search~\cite{PhysRevD.77.032003}. We use the atmospheric neutrino flux from the Honda-Kajita-Kasahara-Midorikawa model in Refs.~\cite{PhysRevD.75.043006,PhysRevD.83.123001}. Neutrino interactions are simulated using \texttt{NEUT}~\cite{HAYATO2002171,NEUT2021}. Among these interactions, charged-current quasielastic scattering (CCQE) and neutral-current single-pion production are dominant background channels for the $p\rightarrow\nu\pi^{+}$ and $n\rightarrow\nu\pi^{0}$ searches, respectively, as discussed in Sec~\ref{section:method}. The CCQE is modeled following Nieves, Simo, and Vacas~\cite{PhysRevC.83.045501} and the two-nucleon knockout process, which effectively enhances the rate of QE-like scattering in oxygen, is implemented according to the model by Gran \textit{et al.}~\cite{PhysRevD.88.113007}. Single-pion production is dominated by baryon-resonance excitation, for which the Rein-Sehgal model~\cite{REIN198179} is used. The NC$\pi^0$ production in oxygen is based on the measurement in Ref.~\cite{NAKAYAMA2005255} and SK NC$\pi^0$ control data samples (see Sec. III in Ref.~\cite{PhysRevD.102.112011}). The momentum and energy of bound nucleons are simulated based on the spectral function and Fermi-gas models. The difference in nuclear models between signal and background MCs is treated as systematic uncertainty, as shown in Sec .~\ref {section:method}. Three-flavor neutrino oscillation is considered based on the mixing parameters of $\sin^{2}\theta_{12}=0.307$, $\sin^{2}\theta_{13}=0.0220$, $\sin^{2}\theta_{23}=0.546$, $\Delta m^{2}_{12}=0.753\times 10^{-5}$\,$\text{eV}^{2}$, and $\Delta m^{2}_{23}=2.453\times 10^{-3}$\,$\text{eV}^{2}$~\cite{10.1093/ptep/ptac097}, and $\delta_{CP}$ is assumed to be zero. A 500-yr exposure of the SK detector for the atmospheric neutrino interactions is simulated and scaled for each SK detector phase.

Nuclear interactions within a nucleus, produced from nucleon decay or atmospheric neutrino interactions, are simulated in \texttt{NEUT} using the cascade model~\cite{HAYATO2002171,NEUT2021} and the Woods-Saxon model~\cite{PhysRev.95.577}. The hadron propagation by pions or nucleons in a nucleus is simulated based on the mean free path of nuclear interactions, which is related to scattering, absorption, charge exchange, and hadron production. Compared to the 2014 SK nucleon decay search, the pion-nuclear interaction model has been significantly updated using $\pi^{\pm}$-nucleus experimental data~\cite{PhysRevD.95.012004}. In this update, pion-nuclear absorption has increased by 40\% at a pion momentum range of 300--600\,MeV/$c$, leading to a significant loss in nucleon decay detection efficiency. The true pion momentum by signal MC is presented, and the cumulative fractions of pion-nuclear interactions as a function of neutral pion momentum are compared in Fig.~\ref{figure:picrs}. The propagation and decay of particles in the SK detector and the responses to Cherenkov photons by PMTs are simulated using the \texttt{GEANT} simulation package~\cite{Brun:1082634} based on the detector calibration parameters for each SK phase. For pion propagation, hadronic interactions with nucleons in water are simulated by \texttt{NEUT}, as well as the nuclear effects within the nucleus. Note that our method treats the pion’s initial prehadronic stage in the nucleus as a fully formed pion, resulting in a conservative overprediction of pion-nucleus scattering.
\begin{table*}[!htbp]
			\caption{\label{table:breakdown}Breakdown of the remaining background event fraction with statistical error [\%] for each event selection and neutrino interaction mode: CCQE, charged-current single-pion (CC1$\pi$), charged-current deep-inelastic (CCDIS), neutral-current single-pion (NC1$\pi$), and neutral-current deep-inelastic (NCDIS). The event fraction is averaged over SK I-V with the corresponding detector's live time.}
	\begin{ruledtabular}
		\begin{tabular}{lccccc}
			& \multicolumn{5}{c}{Neutrino interaction mode} \\
      \cline{2-6}
			Event selection & CCQE & CC1$\pi$ & CCDIS & NC1$\pi$ & NCDIS \\ \hline
			$p\rightarrow\nu\pi^{+}$ (0-decay-$e$) & 72.8$\pm$0.4 & 10.3$\pm$0.1 & 0.8$\pm$0.0 & 13.5$\pm$0.1 & 2.6$\pm$0.1 \\
			$p\rightarrow\nu\pi^{+}$ (1-decay-$e$) & 82.6$\pm$0.2 & 13.2$\pm$0.1 & 1.1$\pm$0.0 & 2.2$\pm$0.0  & 1.0$\pm$0.0 \\
			$n\rightarrow\nu\pi^{0}$ & 5.4$\pm$0.1 & 5.0$\pm$0.1 & 0.8$\pm$0.0 & 80.8$\pm$0.5 & 8.0$\pm$0.1 \\
		\end{tabular}
	\end{ruledtabular}
\end{table*}
\section{SEARCH METHOD}\label{section:method}
The SK I-V phases correspond to 6511 detector live days. The predicted signal purity is enhanced by vetoing the dominant backgrounds using an array of selection cuts caused by cosmic ray muons, low-energy radioactivity, and flashing PMTs~\cite{PhysRevD.71.112005,PhysRevD.102.112011}. Additionally, we require events to be reconstructed within the fiducial volume, to exhibit sufficient visible energy deposit, and to show no activity in the OD. Remaining events are fully contained in the fiducial volume. The same data reduction procedures are applied to signal and background MC events. The nominal MC prediction implies that more than 30\% of the potential signal events are lost in the procedures because the visible energy deposit is suppressed by the pion absorption.

Since the 2014 SK nucleon decay search, the fiducial volume has been expanded by dedicated studies on the reduction of non-neutrino background events and improved event reconstruction near the ID walls~\cite{PhysRevD.102.112011,PhysRevD.109.072014}. The new method reduces the misreconstructed cosmic ray muons and the misidentification between $e$ and $\mu$. The expanded fiducial volume is defined as the region inside the ID located at least 1\,m away from the walls, where the previous analysis used 2\,m for the criterion. The total mass within the fiducial volume corresponds to 27.2\,kt, a 21\% increase compared to the former analysis.

In this analysis, an event reconstruction algorithm, APFit \cite{doi:10.1142/9789819801107_0008,SHIOZAWA1999240}, is used. In APFit, the event vertex is reconstructed by scanning the point where the time-of-flight-corrected PMT timing distribution has the sharpest peak. In the PMT residual times, the timing resolution of the PMT and the track length of the charged particle are taken into account. The number of Cherenkov rings is determined based on a pattern recognition algorithm known as the Hough transformation~\cite{davies1997machine}. Each reconstructed Cherenkov ring is identified as either a showering particle ($e^{\pm},\gamma$) or a nonshowering particle ($\mu^{\pm},\pi^{\pm}$) based on likelihood evaluations with observed PMT hit pattern and expected charge distributions. For single-ring events, the Cherenkov opening angle is additionally considered for the particle identification, and then the event vertex is precisely fitted using the PMT charge information with the Cherenkov opening angle and the particle type. For multiring events, the observed PMT charges are assigned separately for each reconstructed Cherenkov ring. The momentum of each ring is reconstructed from the total PMT charge within a \ang{70} with respect to the ring direction, with corrections for overlapping rings and the direction of incoming Cherenkov light. Michel electrons are identified by detecting PMT hit clusters occurring after the primary event trigger, provided that the number of hits exceeds a threshold and the charge remains below a predefined limit.
\begin{table*}[!htbp]
	\centering
			\caption{\label{table:sysdet} List of detector dependent uncertainties for $p\rightarrow\nu\pi^{+}$ and $n\rightarrow\nu\pi^{0}$ searches. Common uncertainties in event reduction and reconstruction are considered for both signal and background. The first column shows the name of the systematic error, and the next ten columns show the best-fit $\epsilon_j$ in units of $\sigma_{j}$ and $1\sigma$ error size in percent for SK I-V, respectively.}
	\begin{ruledtabular} 
		\renewcommand{\arraystretch}{1.1}
		\begin{tabular}{lcccccccccc}
			\multirow{2}{*}{Systematic uncertainty} & \multicolumn{2}{c}{SK-I} & \multicolumn{2}{c}{SK-II} & \multicolumn{2}{c}{SK-III} & \multicolumn{2}{c}{SK-IV} & \multicolumn{2}{c}{SK-V} \\
			\cline{2-3} \cline{4-5} \cline{6-7} \cline{8-9} \cline{10-11} \rule{0pt}{1em} & Fit value & $\sigma$ & Fit value & $\sigma$ & Fit value & $\sigma$ & Fit value & $\sigma$ & Fit value & $\sigma$ \\
			\hline
      $p\rightarrow\nu\pi^{+}$ search & & & & & & & & & & \\
			FC reduction & 0.008 & 0.2 & $-0.041$ & 0.2 & 0.004 & 0.8 & 0.293 & 1.3 & 0.005 & 1.7 \\
			Non-$\nu$ background ($\mu$-like) & 0.062 & 1 & $-0.073$ & 1 & $-0.007$ & 1 & 0.039 & 1 & 0.058 & 1 \\
			Fiducial volume & 0.082 & 2 & $-0.409$ & 2 & 0.010 & 2 & 0.450 & 2 & 0.006 & 2 \\
			Ring separation & 0.577 & 10 & $-0.196$ & 10 & 0.312 & 10 & $-0.166$ & 10 & $-0.061$ & 10 \\
			Particle identification (one-ring) & 0.040 & 1 & $-0.020$ & 1 & 0.020 & 1 & 0.133 & 1 & 0.028 & 1 \\
			Energy calibration & $-0.340$ & 3.3 & 0.219 & 2.0 & $-0.169$ & 2.4 & $-0.127$ & 2.1 & $-0.579$ & 1.8 \\
			Up and down asymmetry energy calibration & 0.123 & 0.6 & 0.058 & 1.1 & $-0.069$ & 0.6 & 0.062 & 0.5 & $-0.014$ & 0.7 \\
			Decay-$e$ tagging & $-0.923$ & 10 & $-0.464$ & 10 & $-0.363$ & 10 & 0.772 & 10 & 0.412 & 10 \\
			$n\rightarrow\nu\pi^{0}$ search & & & & & & & & & & \\
			FC reduction & $-0.018$ & 0.2 & 0.003 & 0.2 & 0.038 & 0.8 & 0.078 & 1.3 & $-0.066$ & 1.7 \\
			Non-$\nu$ background ($e$-like) & $-0.008$ & 1 & 0.005 & 1 & 0.020 & 1 & 0.001 & 1 & $-0.036$ & 1 \\
			Fiducial volume & $-0.182$ & 2 & 0.032 & 2 & 0.096 & 2 & 0.120 & 2 & $-0.078$ & 2 \\
			Energy calibration & $-0.224$ & 3.3 & $-0.338$ & 2.0 & 0.117 & 2.4 & 0.275 & 2.1 & $-0.055$ & 1.8 \\
			Up and down asymmetry energy calibration & $-0.082$ & 0.6 & $-0.098$ & 1.1 & $-0.041$ & 0.6 & $-0.106$ & 0.5 & 0.005 & 0.7 \\
			Sub-GeV two-ring $\pi^{0}$ selection & $-0.593$ & 7.1 & 0.067 & 4.3 & 0.086 & 1.6 & 0.339 & 5.4 & $-0.104$ & 2.5 \\
			Decay-$e$ tagging & 0.019 & 10 & $-0.003$ & 10 & $-0.010$ & 10 & $-0.008$ & 10 & 0.005 & 10 \\
		\end{tabular}
	\end{ruledtabular}
\end{table*}
\begin{table*}[!htbp]
\centering
\caption{\label{table:sysphys} List of physics model uncertainties. For the signal, uncertainties in the Fermi momentum and correlated nucleon decay are considered. For background, uncertainties in neutrino interactions, pion production, atmospheric neutrino flux, and neutrino oscillation are considered. For both signal and background, uncertainties in pion-nuclear interactions are considered. The first column shows the name of the systematic error, the next two columns show the best-fit $\epsilon_{j}$ in units of $\sigma_{j}$, and the last column shows the $1\sigma$ error size in percent.}
\begin{ruledtabular} 
\renewcommand{\arraystretch}{1.1}
\begin{tabular}{llccc}
\multicolumn{2}{c}{\multirow{2}{*}{Systematic uncertainty}} & \multicolumn{2}{c}{Fit value} & \multirow{2}{*}{$\sigma$} \\
\cline{3-4} \rule{0pt}{1em} & & $p\rightarrow\nu\pi^{+}$ & $n\rightarrow\nu\pi^{0}$ & \\ \hline
Fermi momentum & & 0.000 & 0.000 & 10 \\
Correlated nucleon decay & & 0.000 & 0.000 & 100 \\
\multirow{2}{*}{Pion FSI and SI} & Min & $-0.218$ & 0.833 & 10 \\
& Max & 0.046 & 0.294 & 10 \\
Single $\pi$ production, axial coupling & & 0.363 & $-0.272$ & 10 \\
Single $\pi$ production, C$_{\mathrm{A5}}$ & & 0.355 & 0.475 & 10 \\
Single $\pi$ production, background & & 0.197 & $-0.097$ & 10 \\
Single $\pi$ production, $\pi^{0}/\pi^{\pm}$ ratio & & $-0.248$ & 0.564 & 40 \\
Single $\pi$ production, $\bar\nu/\nu$ ratio & & 0.178 & 0.170 & 10 \\
Coherent $\pi$ production & & 0.113 & $-0.181$ & 100 \\
M$_A$ in QE & & 0.733 & $-0.055$ & 10 \\
CCQE cross section, shape & & 0.811 & $-0.018$ & 10 \\
\multirow{2}{*}{CCQE cross section, normalization} & $E_{\nu} < 1.33$ GeV  & $-0.059$ & 0.006 & 10 \\
& $E_{\nu} > 1.33$ GeV & 0.108 & $-0.012$ & 10 \\
CCQE cross section, $\bar\nu/\nu$ ratio & & $-0.016$ & $-0.015$ & 10 \\
CCQE cross section, $\mu/e$ ratio & & 0.171 & 0.002 & 10 \\
Meson exchange current & & $-0.008$ & $-0.092$ & 10 \\
NC/CC ratio & & 0.310 & 0.065 & 20 \\
NC fraction from hadron simulation & & 0.155 & & 10 \\
DIS cross section & & $-0.025$ & $-0.024$ & 10 \\
DIS model difference & & $-0.141$ & $-0.143$ & 10 \\
DIS $Q^{2}$ distribution ($W>2$ GeV/c$^2$) & & $-0.012$ & $-0.020$ & 10 \\
\multirow{3}{*}{DIS $Q^{2}$ distribution ($W<2$ GeV/c$^2$)} & Vector & 0.011 & 0.007 & 10 \\
& Axial & 0.017 & 0.020 & 10 \\
& Normalization & $-0.003$ & $-0.002$ & 10 \\
DIS hadron multiplicity & & 0.001 & $-0.009$ & 10 \\
\multirow{2}{*}{Flux normalization} & $E_{\nu}<1$ GeV & $-0.730$ & 0.180 & 25 \\
& $E_{\nu}>1$ GeV & 0.170 & $-0.117$ & 15 \\
\multirow{3}{*}{Flux, $(\nu_\mu+\bar\nu_\mu)$/$(\nu_e+\bar\nu_e)$ ratio} & $E_{\nu}<1$ GeV & 0.021 & 0.009 & 2 \\
& $1<E_{\nu}<10$ GeV & 0.031 & 0.000 & 3 \\
& $E_{\nu}>10$ GeV & $-0.001$ & 0.003 & 5 \\
\multirow{3}{*}{Flux, $\bar\nu_e$/$\nu_e$ ratio} & $E_{\nu}<1$ GeV & 0.028 & 0.002 & 5 \\
& $1<E_{\nu}<10$ GeV & 0.004 & $-0.015$ & 5 \\
& $E_{\nu}>10$ GeV & $-0.000$ & 0.002 & 8 \\
\multirow{3}{*}{Flux, $\bar\nu_\mu$/$\nu_\mu$ ratio} & $E_{\nu}<1$ GeV & $-0.052$ & 0.002 & 2 \\
& $1<E_{\nu}<10$ GeV & $-0.025$ & $-0.025$ & 6 \\
& $E_{\nu}>10$ GeV & $-0.000$ & 0.001 & 6 \\
Flux, up/down ratio & & $-0.068$ & $-0.005$ & 1 \\
Flux, horizontal/vertical ratio & & $-0.035$ & $-0.002$ & 1 \\
$K$/$\pi$ ratio in flux calculation & & 0.033 & $-0.005$ & 10 \\
Neutrino path length & & $-0.045$ & 0.001 & 10 \\
\multirow{5}{*}{Solar activity} & SK-I & 0.008 & $-0.059$ & 20 \\
& SK-II & $-0.373$ & 0.051 & 50 \\
& SK-III & 0.013 & 0.014 & 20 \\
& SK-IV & 0.032 & 0.012 & 7 \\
& SK-V & 0.018 & $-0.014$ & 20 \\
$\Delta{m}^{2}_{21}$ & & 0.023 & 0.000 & 0.00018 \\
$\mathrm{sin}^{2}(\theta_{12})$ & & 0.026 & 0.001 & 1.3 \\
$\mathrm{sin}^{2}(\theta_{13})$ & & 0.001 & 0.000 & 0.07 \\
Matter effects & & $-0.003$ & $-0.002$ & 6.8 \\
\end{tabular}
\end{ruledtabular}
\end{table*}
\subsection{Event selection}
The event selection criteria (C1--C5) for the search for $p\rightarrow\nu\pi^{+}$ and $n\rightarrow\nu\pi^{0}$ are as follows based on the event reconstruction:
\begin{itemize}[noitemsep]
\item[C1] one Cherenkov ring for $p\rightarrow\nu\pi^{+}$ and two Cherenkov rings for $n\rightarrow\nu\pi^{0}$;
\item[C2] particle identification (PID) with a nonshowering ring for $p\rightarrow\nu\pi^{+}$ and all showering rings for $n\rightarrow\nu\pi^{0}$;
\item[C3] zero or one Michel electron for $p\rightarrow\nu\pi^{+}$ and no Michel electron for $n\rightarrow\nu\pi^{0}$;
\item[C4] reconstructed mass $ M_{\mathrm{tot}}$ should satisfy $85 < M_{\mathrm{tot}} < 185$ MeV/$c^{2}$ for $n\rightarrow\nu\pi^{0}$;
\item[C5] reconstructed total momentum $P_{\mathrm{tot}}$ should satisfy and $200 \le P_{\mathrm{tot}} < 1000$ MeV/$c$ for $p\rightarrow\nu\pi^{+}$ and $0 < P_{\mathrm{tot}} < 1000$ MeV/$c$ for $n\rightarrow\nu\pi^{0}$.
\end{itemize}
From C1 to C2, the number of rings and corresponding particle types are considered according to the event topology of the signal for each mode. Since the neutrino is invisible to the detector, the nucleon decay signal is traced by a single pion event: a nonshowering Cherenkov ring from a single $\pi^{+}$ produced in $p\rightarrow\nu\pi^{+}$, and by two showering Cherenkov rings from $\pi^{0}\rightarrow\gamma\gamma$ for $n\rightarrow\nu\pi^{0}$. In C3, the number of Michel electrons required is based on the number of (anti)muons in the signal. For both modes, events without the Michel electron are allowed. For $p\rightarrow\nu\pi^{+}$, one Michel electron is also allowed to cover the signal events with $\pi^{+}\rightarrow\mu^{+}\nu$ instead of $\pi^{+}$ hadronic absorption by water. The final event samples with different numbers of Michel electrons are treated independently. In C4 and C5, using reconstructed momentum and PID for each ring, total mass $M_{\mathrm{tot}}$ and total momentum $P_{\mathrm{tot}}$ are evaluated as follows:
\begin{align}
	P_{\mathrm{tot}} = & \left| \sum_{i} \vec{p}_i \right| \\
	E_{\mathrm{tot}} = & \sum_{i} \left| \vec{p}_i \right| \\
	M_{\mathrm{tot}} = & \sqrt{E_{\mathrm{tot}}^2-P_{\mathrm{tot}}^2}
\end{align}
where the summation is over the number of rings and $\vec{p}_i$ is the reconstructed momentum of $i$th ring. In C4, the total mass cut around the physical mass of $\pi^{0}$ is considered for $n\rightarrow\nu\pi^{0}$. After C5, the range of the total momentum is restricted below 1000\,MeV/$c$, which sufficiently covers the signal momentum range and the tail of the background distribution constrains its overall normalization. For $p\rightarrow\nu\pi^{+}$, an additional lower limit by 200\,MeV/$c$ is applied, vetoing events below the Cherenkov threshold by nonshowering particles. The breakdown of remaining background samples after event selections is shown in Table~\ref{table:breakdown}. For the $p\rightarrow\nu\pi^{+}$ search, the dominant background events originate from a single nonshowering ring by $\overset{\scriptscriptstyle(-)}{\nu_{\mu}}$ CCQE interactions. As muon neutrinos undergo approximately 50\% disappearance in the energy range relevant to this search, incorporating neutrino oscillations in the MC reduces the expected atmospheric neutrino background relative to an unoscillated flux. The next dominant background events arise from single-pion production (1$\pi$). Among single-pion events, the event sample with zero decay electrons has a larger fraction of neutral-current single-pion production (NC1$\pi$) than the one-decay-electron sample due to the restriction of secondary decay by the (anti)muon after CC1$\pi$. NC1$\pi$ is the dominant background in the $n\rightarrow\nu\pi^{0}$ search.

Table~\ref{table:bestfit} shows the expected signal efficiencies. Compared to the 2014 SK nucleon decay search, the averaged signal efficiencies are decreased by 16\% for $p\rightarrow\nu\pi^{+}$ and by 32\% for $n\rightarrow\nu\pi^{0}$ due to the updated pion-nuclear interaction model, which increased pion absorption rates for both modes. For the $p\rightarrow\nu\pi^{+}$ search, efficiencies for the one-decay-electron sample have increased since SK-IV with improved Michel-electron-tagging efficiency, which reduced the inefficiency in the zero-decay-electron sample.
\subsection{Spectrum analysis}
The $p\rightarrow\nu\pi^{+}$ and $n\rightarrow\nu\pi^{0}$ searches expect approximately 13000\, and 2000\, background events, respectively for SK I-V periods in total. In each case, this is expected to have a relative presence 107 and 55 times greater than nucleon decay signals assuming the lifetime limits from the 2014 SK nucleon decay search. To effectively discriminate between signal and background, we search for a signal bump above the data spectrum defined by reconstructed total momentum as in the 2014 SK nucleon decay search. For $p\rightarrow\nu\pi^{+}$, momentum reconstruction is based on the muon hypothesis instead of the charged pion due to the dominance of single $\mu$ from CCQE in the background events and better momentum resolution of muon hypothesis compared to charged pion. For both modes, the momentum bin width is 50 MeV/$c$; a total of 160 bins are used for the $p\rightarrow\nu\pi^{+}$ search (16 momentum bins $\times$ 5 SK periods $\times$ 2 event samples) and 100 bins are used for the $n\rightarrow\nu\pi^{0}$ search (20 momentum bins $\times$ 5 SK periods). These momentum bins are simultaneously fitted for each nucleon decay mode. The spectral fit is conducted by performing a $\chi^{2}$ minimization over the parameter space defined by global scale factor $\beta$ for signal normalization. The $\chi^{2}$ statistic is based on the Poisson probability for each momentum bin content and quadratic penalty terms, which account for systematic errors. The $\chi^2$ is defined as
\begin{align}
	\chi^{2} = 2 \sum_{i}\left(E_{i}+O_{i}\left[\ln\frac{O_{i}}{E_{i}}-1\right]\right)+\sum_{j}\left(\frac{\epsilon_{j}}{\sigma_{j}}\right)^{2} \label{eq:chi2}\\ 
	E_{i} = \left[E_{i}^{\text{bkg}}+\beta E_{i}^{\text{sig}}\right]\left(1+\sum_{j}f_{ij}\epsilon_{j}\right) \label{eq:expectation}
\end{align}
\noindent where the index $i$ is each momentum bin, $j$ is index of systematic error, $O_i$ is the number of observed events, and $E_i$ is the expected number of events at given $\beta$, the product of a nominal expectation by signal and background MCs and a scale factor. The $\sigma_j$ is the size of the $j$th systematic error, and $f_{ij}$ is the fractional change of the $i$th bin content by $\sigma_j$ of the $j$th systematic error. With obtained $f_{ij}$ and known $\sigma_j$, a nuisance parameter $\epsilon_j$ is fitted for each systematic error by solving the equations $\partial\chi^{2}/\partial\epsilon_j=0$.
\begin{figure*}[!htbp]
	\includegraphics[width=0.95\textwidth]{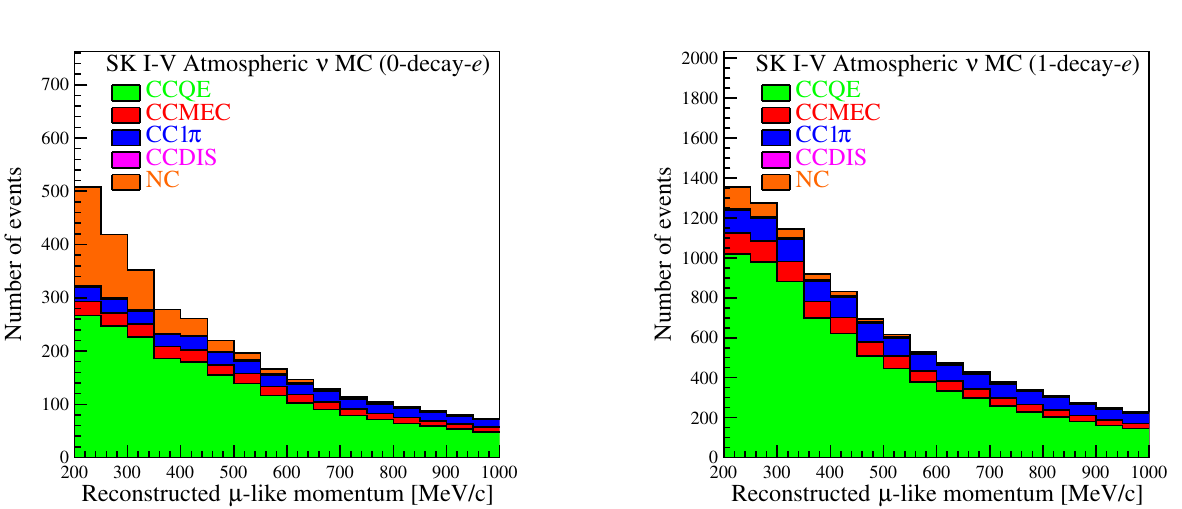}
	\caption{\label{figure:background}Expected breakdown of background spectra for $p\rightarrow\nu\pi^{+}$ search with zero-decay-electron sample (left) and one-decay-electron sample (right). The breakdown by neutrino interactions includes charged-current quasielastic scattering with one-nucleon knockout (green) and two-nucleon knockout (red) by meson exchange current, single-$\pi$ production (blue), deep-inelastic scattering (magenta), and neutral-current interactions (orange). The MC event rate for each SK phase is scaled based on the SK I-V live time, respectively.}
\end{figure*}
\begin{figure*}[!hbtp]
	\centering
	\includegraphics[width=0.90\textwidth]{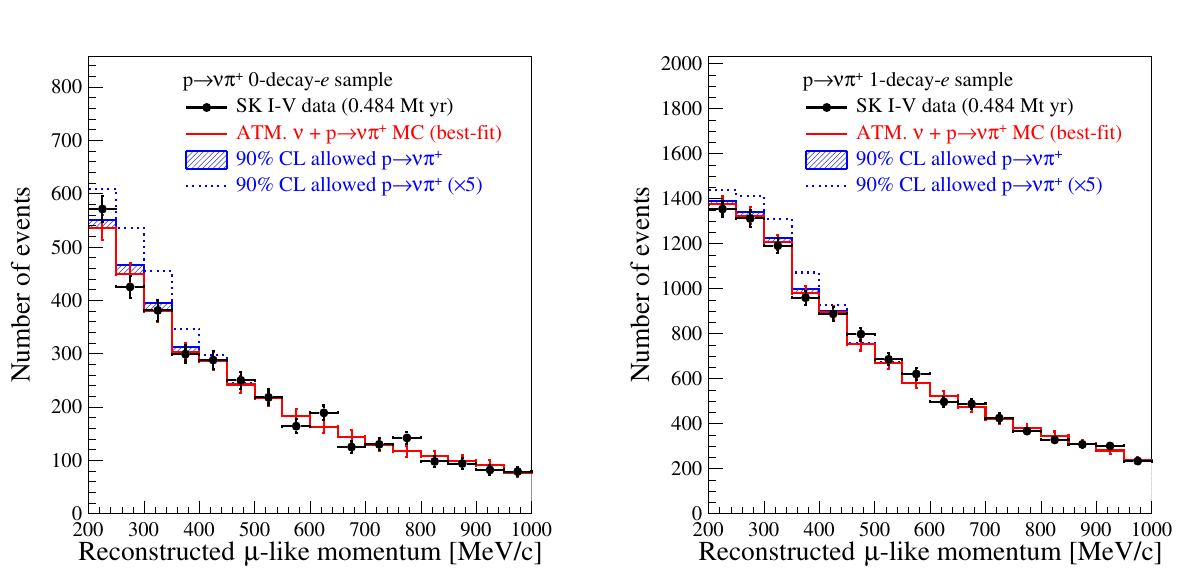}
	\caption{\label{figure:momentum1} Reconstructed momentum distributions for SK data (black dots), the best fit for atmospheric neutrino background and $p\rightarrow\nu\pi^{+}$ nucleon decay Monte Carlo simulation (red histogram), 90\% confidence level allowed amount of nucleon decay (blue hatched box). The dashed blue line shows how a positive signal of nucleon decay would look, corresponding to 5 times the limit we set on the decay partial lifetimes. The left panel shows 0-decay-$e$ sample, and the right panel shows 1-decay-$e$ sample.}
\end{figure*}
\subsection{Systematic uncertainties}
We consider additional uncertainties in physics models and detector systematics compared to the 2014 SK nucleon decay search. For nucleon decay, the uncertainty in the Fermi momentum is estimated using the ratio of nucleon momentum distributions from different models used in the atmospheric neutrino interactions, and the uncertainty in the correlated nucleon decay is set to 100\% as in the former analysis~\cite{PhysRevLett.115.121803}. For background-specific errors, we include contributions from atmospheric neutrino flux, neutrino interactions, and neutrino oscillation. Flux-related errors are evaluated based on uncertainties in the hadronic interactions and air density \cite{PhysRevD.75.043006}, and comparison between the Honda-Kajita-Kasahara-Midorikawa model with others \cite{PhysRevD.70.023006,BATTISTONI2003269,BATTISTONI2003291}. Errors by the CCQE cross section are evaluated by comparing the Fermi-gas models \cite{SMITH1972605,PhysRevC.83.045501} and by assigning an uncertainty to the axial-form-factor parameter $M_{\mathrm{A}}^{\mathrm{QE}}=(1.05\pm0.16)$ GeV/$c^{2}$. For single-pion production, uncertainties in the Rein-Sehgal model \cite{REIN198179} and its comparison with the Hernandez model \cite{PhysRevD.76.033005} are mainly considered. For NC events with hadron production, the uncertainty for contamination of charged-pion events with $\mu$-like events is set by 10\%. For both signal and background, uncertainties in nuclear effects on pion final state interaction (FSI) and secondary interaction (SI) are considered using 16 sets of the interaction probabilities in the \texttt{NEUT} cascade model which are representative for $1\sigma$ errors from a fit to external pion scattering data \cite{PhysRevD.91.072010}. Among 16 sets, two conservative sets are chosen based on the variation of signal events, which give maximal and minimal interaction rates around the range of signal pion momentum. Uncertainties related to the physics models are common in each SK phase, except for the solar activity, which deals with the time variation of solar wind and its impact on atmospheric neutrino flux. The detector-dependent uncertainties related to performance in event reduction, event reconstruction, and calibrations are considered independent error sources for each SK phase. Details of physics model uncertainties for atmospheric neutrino and detector-dependent errors are described in Refs.~\cite{PhysRevD.97.072001,PhysRevD.109.072014}.

The full lists of systematic errors are summarized in Table~\ref{table:sysdet} and~\ref{table:sysphys}. Among them, nuisance parameters $\epsilon_j$ for atmospheric neutrino flux and cross sections of neutrino interactions strongly affect the overall normalization of the background spectrum in the fit. To avoid redundancy with these parameters, we do not fit the overall background normalization, which was considered as $\chi^{2}$ parameter $\alpha$ in the previous analysis. Instead, we fix $\alpha$ as 1 in this analysis and account for its systematic uncertainty through the relevant $\epsilon$ parameters in the fit.
\begin{figure}[!htbp]
	\centering
	\includegraphics[width=0.5\textwidth]{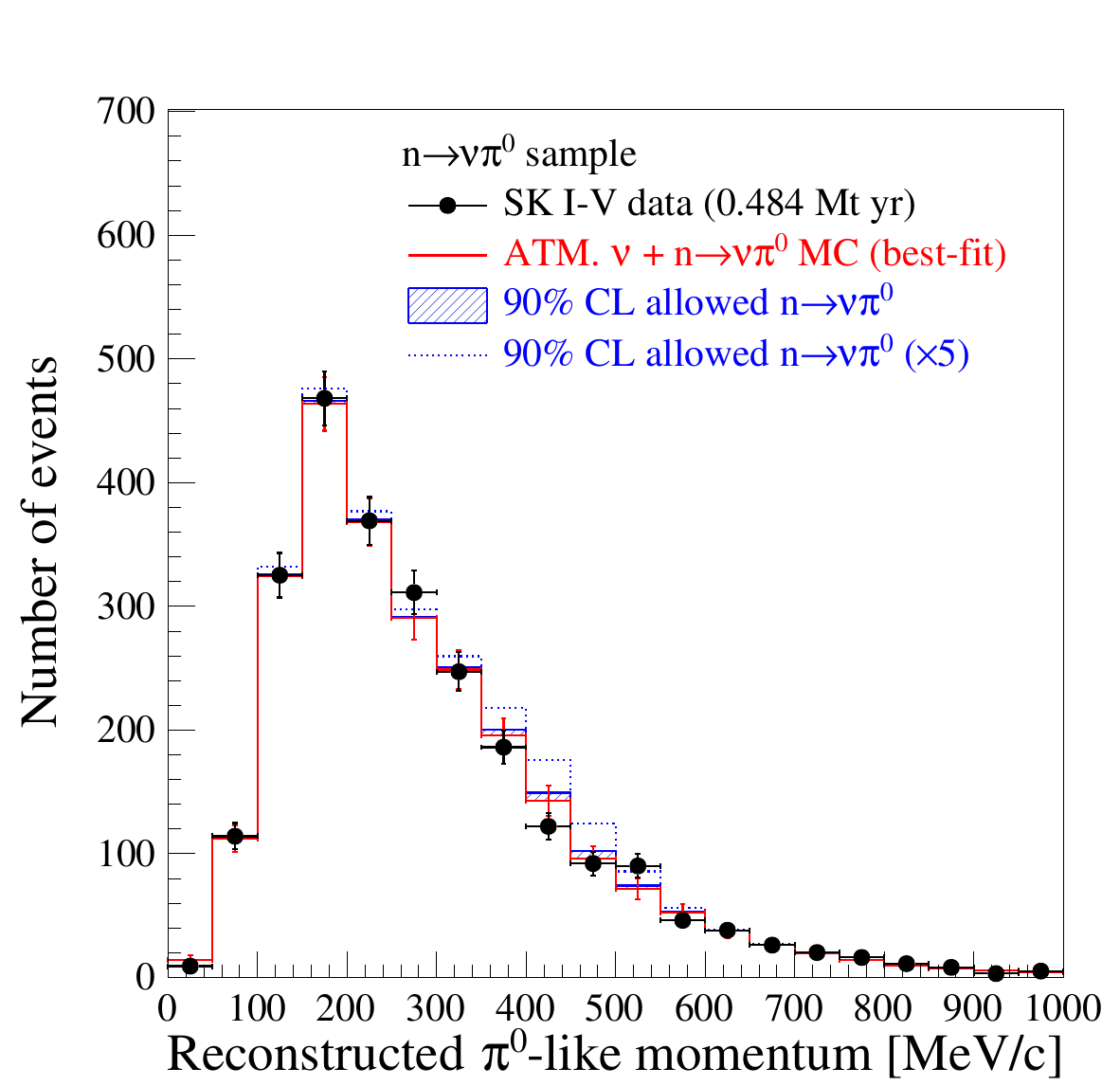}
	\caption{\label{figure:momentum2} Reconstructed momentum distribution for SK data (black dots), the best fit for atmospheric neutrino background and $n\rightarrow\nu\pi^{0}$ nucleon decay Monte Carlo simulation (red histogram), and 90\% confidence level allowed amount of nucleon decay (blue hatched box). The dashed blue histogram shows 5 times the limit we set on the decay partial lifetimes.}
\end{figure}
\begin{table*}[!htbp]
	\caption{\label{table:bestfit}Best-fit parameter values, signal-detection efficiency for each SK period, and lower limits on partial lifetime for each nucleon decay mode at 90\% confidence level.}
	\begin{ruledtabular} 
		\begin{tabular}{ccccccccc}
			\multirow{2}{*}{Decay mode}&&\multicolumn{5}{c}{Signal detection efficiency [\%]}&\multirow{2}{*}{Best-fit $\beta$}&\multirow{2}{*}{$\tau/B$ ($\times10^{32}$\,yr)}\\
			&&SK-I&SK-II&SK-III&SK-IV&SK-V&&\\ \hline
			\multirow{2}{*}{$p\rightarrow\nu\pi^{+}$}&(0-decay-$e$)&$14.5\pm0.1$&$14.2\pm0.1$&$14.5\pm0.1$&$11.6\pm0.1$&$11.7\pm0.1$&\multirow{2}{*}{0.0}&\multirow{2}{*}{3.5}\\
			&(1-decay-$e$)&$15.9\pm0.1$&$15.3\pm0.1$&$15.9\pm0.1$&$18.1\pm0.2$&$18.1\pm0.2$&&\\
			$n\rightarrow\nu\pi^{0}$&&$32.2\pm0.2$&$30.1\pm0.2$&$31.9\pm0.2$&$32.6\pm0.2$&$32.9\pm0.2$&0.0&14.0\\
		\end{tabular}
	\end{ruledtabular}
\end{table*}
\subsection{Sensitivity}
The sensitivities are computed as $\beta_{90\mathrm{CL}}$, which is allowed signal normalization at 90\% confidence level (CL), by using pseudodata constructed based on the background-only hypothesis. Adding the SK IV-V data to the SK I-III data improves the sensitivity by 60\% for both nucleon decay modes. Enlarging the fiducial volume improves the search sensitivity by 5\% for $n\rightarrow\nu\pi^{0}$ and by 10\% for $p\rightarrow\nu\pi^{+}$. However, a new set of systematic uncertainties reduces the sensitivities. For $p\rightarrow\nu\pi^{+}$, the sensitivity is reduced by 65\% with significant contributions from physics model errors associated with the pion-nuclear interactions (44\%), single-pion production (35\%), and NC (44\%), where the spectral contamination is located around the expected signal range, as illustrated in Fig. \ref{figure:background}. For $n\rightarrow\nu\pi^{0}$, the sensitivity is reduced by 40\%, primarily due to model uncertainties in the single-pion production by neutrino interactions (33\%). Final expected sensitivities from SK I-V nominal MCs with expanded fiducial volume are $1.4\times10^{32}$\,yr for $p\rightarrow\nu\pi^{+}$ and $5.3\times10^{32}$\,yr for $n\rightarrow\nu\pi^{0}$, respectively.
\section{SEARCH RESULTS}\label{section:results}
The fit to the data spectrum by reconstructed momentum is performed for SK I--V periods simultaneously with the $\beta$ parameter. The $\chi^{2}$ is computed over the fit parameter space defined by non-negative signal normalization. Figures~\ref{figure:momentum1} and ~\ref{figure:momentum2} show the resulting spectra after the fit with combined SK I-V data. The effect of the systematic errors is included by fitting $\epsilon_j$ with the bin-by-bin response of $f_{ij}$. Summaries of fitted $\epsilon_j$ are listed in Table~\ref{table:sysdet} and~\ref{table:sysphys}. Overall, the systematic pulls ($=\epsilon_j/\sigma_j$) are within $\pm 1$, which implies no strong tension between the best-fit MC and data. For both $p\rightarrow\nu\pi^{+}$ and $n\rightarrow\nu\pi^{0}$ searches, the best fit gives $\beta=0$ with $\chi^2/\nu=178.6/159$ and $77.3/99$ respectively. At the best fit of $p\rightarrow\nu\pi^{+}$ search, dominant systematic pulls come from decay electron tagging and the model uncertainties in CCQE and neutrino flux. For the $n\rightarrow\nu\pi^{0}$ search, the model uncertainties in pion production and its nuclear interactions have comparable size among best-fit pulls. With fitted pulls, the best-fit MC and data spectra show no discrepancy between them for each nucleon decay search. We find no statistically significant indication of nucleon decay signal in the SK I-V data. Therefore, the 90\% CL allowed signal events are determined by the $\Delta\chi^{2}(=\chi^{2}-\chi^{2}_{\mathrm{best}})$ contour over the $\beta$ parameter. Since the fit parameter $\beta$ is constrained to the physical region, i.e., $\beta \geq 0$, critical values for 90\% CL are estimated by the Feldman-Cousins method~\cite{PhysRevD.57.3873,Acero_2025}. For both nucleon decay modes, critical values near $\beta=0$ are smaller than the standard value 2.706 due to the positive-$\beta$ fit constraint. The values converge over the standard value for $p\rightarrow\nu\pi^{+}$ due to systematic uncertainties (e.g., NC related), resulting in spectral shapes similar to that of the signal MC, reducing the $\chi^2$. The partial lifetime limits are then calculated by
\begin{equation}
\tau/B = \frac{\lambda\epsilon N}{N_{90\mathrm{CL}}},
\end{equation} 
where $\lambda$ is detector exposure, $\epsilon$ is signal efficiency, $N$ is the number of source nucleons per kton, and $N_{90\mathrm{CL}}$ is the number of signal events allowed by 90\% CL. The summary of fit results and corresponding partial lifetime limits are shown in Table~\ref{table:bestfit}.
\section{CONCLUSION}\label{section:conclusion}
Searches for nucleon decays via $p\rightarrow\nu\pi^{+}$ and $n\rightarrow\nu\pi^{0}$ are conducted with 0.484\,Mt$\cdot$yr of SK I-V data. Since the 2014 SK nucleon decay search,  physics models for pion-nuclear interactions and pion production by neutrino interactions have been tuned to external data. Systematic uncertainties in physics models related to Fermi momentum, correlated nucleon decay, pion-nuclear interactions, and atmospheric neutrinos are included. The fiducial volume is increased with an improved event reconstruction method. No significant excess of data is found in the expected signal regions. Accordingly, lower bounds on nucleon partial lifetimes are set by $3.5\times10^{32}$\,yr for $p\rightarrow\nu\pi^{+}$ and $1.4\times10^{33}$\,yr for $n\rightarrow\nu\pi^{0}$ at 90\% CL. Against the significant increase of data and the fiducial volume, the new limit for $p\rightarrow\nu\pi^{+}$ is lower, i.e., less constraining, than the 2014 SK results due to the reduced signal efficiencies by the updated pion-nuclear interaction model and rigorous estimation of systematic uncertainties. In contrast, the new limit for $n\rightarrow\nu\pi^{0}$ is more stringent than the previous value of $1.1\times10^{33}$\,yr. The new results will offer more robust constraints on viable GUT models. We anticipate examining these nucleon decay modes by next-generation experiments such as Hyper-Kamiokande and the Deep Underground Neutrino Experiment through their substantially larger detector exposures and with improved control of systematic uncertainties.
\begin{acknowledgments}
We gratefully acknowledge the cooperation of the Kamioka Mining and Smelting Company. The Super-Kamiokande experiment has been built and operated from funding by the Japanese Ministry of Education, Culture, Sports, Science and Technology; the U.S. Department of Energy; and the U.S. National Science Foundation. Some of us have been supported by funds from the National Research Foundation of Korea (NRF-2009-0083526, NRF-2022R1A5A1030700, NRF-2022R1A3B1078756, RS-2025-00514948) funded by the Ministry of Science, Information and Communication Technology (ICT); the Institute for Basic Science (IBS-R016-Y2); and the Ministry of Education (2018R1D1A1B07049158, 2021R1I1A1A01042256, RS-2024-00442775); the Japan Society for the Promotion of Science; the National Natural Science Foundation of China (Grants No. 12375100 and No. 12521007); the Spanish Ministry of Science, Universities and Innovation (Grant No. PID2021-124050NB-C31); the Natural Sciences and Engineering Research Council (NSERC) of Canada; the Scinet and Digital Research of Alliance Canada; the National Science Centre (UMO-2018/30/E/ST2/00441 and UMO-2022/46/E/ST2/00336) and the Ministry of  Science and Higher Education (2023/WK/04), Poland; the Science and Technology Facilities Council (STFC) and Grid for Particle Physics (GridPP), United Kingdom; the European Union’s Horizon 2020 Research and Innovation Programme H2020-MSCA-RISE-2018 JENNIFER2 Grant Agreement No. 822070, H2020-MSCA-RISE-2019 SK2HK Grant Agreement No. 872549; European Union's Next Generation EU/PRTR  Grant No. CA3/RSUE2021-00559; and the National Institute for Nuclear Physics (INFN), Italy.

\end{acknowledgments}
\section*{Data availabitiy}
The data that support the findings of this article are openly available~\cite{jung_2026_18242872}.
\bibliography{skndk2025}
\end{document}